# Tuning the vertical location of helical surface states in topological insulator heterostructures via dual-proximity effects

Guangfen Wu[1,2], Hua Chen[3,4], Yan Sun[5], Xiaoguang Li[6,3,1], Ping Cui[1,3], Cesare Franchini[5], Jinlan Wang[2], Xing-Qiu Chen[5] and Zhenyu Zhang[1,7,4]

[1]ICQD, Hefei National Laboratory for Physical Sciences at the Microscale, University of Science and Technology of China, Hefei, Anhui, 230026, China

[2]Department of Physics, Southeast University, Nanjing, 211189, China

[3]Department of Physics, University of Tennessee, Knoxville, TN 37996, USA

[4]Department of Physics, University of Texas, Austin, TX 78712, USA

[5]Shenyang National Laboratory for Materials Science, Institute of Metal Research, Chinese Academy of Sciences, Shenyang, 110016, China

[6]State Key Laboratory of Surface Physics and Department of Physics, Fudan University, Shanghai, 200433, China

[7]School of Applied Science and Engineering, Harvard University, Cambridge, MA 02138, USA

**In integrating topological insulators (TIs) with conventional materials, one crucial issue is how the topological surface states (TSS) will behave in such heterostructures. We use first-principles approaches to establish accurate tunability of the vertical location of the TSS via intriguing dual-proximity effects. By depositing a conventional insulator (CI) overlayer onto a TI substrate ($Bi_2Se_3$ or $Bi_2Te_3$), we demonstrate that, the TSS can float to the top of the CI film, or stay put at the CI/TI interface, or be pushed down deeper into the otherwise structurally homogeneous TI**

**substrate. These contrasting behaviors imply a rich variety of possible quantum phase transitions in the hybrid systems, dictated by key material-specific properties of the CI. These discoveries lay the foundation for accurate manipulation of the real space properties of TSS in TI heterostructures of diverse technological significance.**

The discoveries of topological insulator (TI) phases[1-4] in HgTe quantum wells[5] and Bi-based materials ($Bi_{1-x}Sb_x$, $Bi_2Se_3$, $Bi_2Te_3$, etc.)[6-13] have inspired extensive research efforts on exploiting their exotic quantum properties [14-18] and potential applications in e.g. quantum computing[19], spintronics[20], and catalysis[21,22]. The uniqueness of TI lies in the robust metallic topological surface states (TSS) with linear Dirac-cone-like energy-momentum dispersion. The TSS arises from the strong intrinsic spin-orbit coupling (SOC) that drives the electronic structures of the TIs to a topologically nontrivial phase, and is expected to appear at an interface between topologically nonequivalent regions, such as between a TI and a conventional insulator (CI). The requirement of strong SOC suggests that heavy-element, small-band-gap semiconductors are the most promising candidate materials[11-13,23] to reach the TI phase. Recent research efforts have also revealed that a trivial insulator can be twisted into a topological state by manipulating the spin-orbit interaction[24,25], the crystal lattice[11,13], or with the application of an external electric field[26], driving the system through a topological phase transition.

The present study offers a conceptually intriguing alternative approach to achieving such quantum phase transitions, via dual topological proximity effects, characterized by controlled shift of the spatial location of the TSS in TI-based heterostructures. Recently, a number of peculiar quantum phenomena have been discovered in various TI-based heterostructures[14-21,27-35], such as Majorana fermions induced by the superconducting proximity effect[19], topological magnetoelectric effect[4], quantized anomalous Hall effect[14], and electron reservoir effect of TSS in surface catalysis[21]. Although the TSS was confirmed to be robust as the TI surface is intruded under diverse conditions provided that time reversal symmetry is preserved[21, 27-31], its properties can be largely modified due to the tunability in the magnitude of interfacial coupling within the heterostructures.

Specifically, the present work reveals a new degree of freedom in the tunability of the TSS properties, namely, its location perpendicular to the surface, as a CI overlayer is placed onto a TI substrate. We establish this intriguing and highly desirable tunability using ultrathin (single stoichiometric layer) films of CIs (Zn$M$, $M$ = S, Se, and Te) on Bi$_2$Se$_3$ or Bi$_2$Te$_3$ as prototypical examples, and the concepts established are expected to be applicable to other related heterostructural systems or when thicker overlayers are used. We show that, because of the strong influence of the interaction with the TI on the electronic properties of the ultrathin CI films, the latter may undergo a nontrivial topological phase transition, signified by the relocation of the TSS from the TI/CI interface to the top of the CI film (Fig. 1a). Furthermore, the TSS can also be manipulated to stay put at the TI/CI interface (Fig. 1b) by tuning one or more key physical properties of the CI to be out of the window for topological phase transition. Most strikingly, we discover a reverse-proximity effect: The top quintuple layer (QL) of the TI may even be driven by the CI film to a topologically trivial phase, signified by a spatial shift of the TSS toward the inside of the TI (Fig. 1c). Collectively, these results help to enrich our understanding of possible topological quantum phase transitions in the hybrid systems, and also lay the foundation for precise manipulation of the real space properties of TSS in TI heterostructures of diverse technological significance. Here we also wish to emphasize one important aspect associated with the introduction of "topological proximity effect": It does not involve symmetry breaking associated with a local order parameter, as the topological phase is a nonlocal, global property of the system[36]. It is qualitatively different from traditional proximity effects[19,37,38], which typically invoke some symmetry breaking processes measured by the corresponding local order parameters.

**Results and discussions**

A bulk Zn$M$ ($M$ = S, Se, and Te) material crystallizes in a zincblende structure (space group $Fd$-$3m$), with its (111) surface geometry matching well with the (0001) surface of Bi$_2$Se$_3$ or Bi$_2$Te$_3$. The lattice mismatch between Zn$M$ (111) and Bi$_2$Se$_3$ (0001) is +8.4%, +3.5%, and -3.9% for $M$ = S, Se, and Te, respectively, where "+" or "-" denotes tensile or compressive strain. Another advantage of using the Zn$M$ family is that they span a considerable range of the key property parameters such as the SOC strength and band gap[34] (see Table I, and Supplementary Section S1). As proof of principles, here we limit ourselves to the cases where a single stoichiometric Zn$M$ layer is deposited onto a TI substrate (Fig. 1d).

**Topological proximity effects in ZnS/Bi$_2$Se$_3$ and ZnSe/Bi$_2$Se$_3$.** Figures 2a-c display the atom-resolved band structures of ZnS/Bi$_2$Se$_3$. The Dirac points of the TSSs are still located at the Fermi level, indicating a negligible charge transfer (or transfer doping[39]) between the ZnS film and Bi$_2$Se$_3$. Although the existence of the Dirac points is guaranteed by time reversal symmetry, away from the $\Gamma$ point, the degeneracy of the two TSSs from the two surfaces of the TI slab is lifted, due to the interaction between the ZnS film and the upper surface of the Bi$_2$Se$_3$. To find out the spatial location of the TSS, we plot the atom-specific character of each band, as indicated by the dots superposed onto the band structure. It is apparent from Figs. 2b and 2c that the Dirac-cone like TSS bands are localized on the 1$^{st}$ QL (upper surface) and 6$^{th}$ QL (lower surface) of the Bi$_2$Se$_3$, whereas from Fig. 2a we see that the TSS has negligible electron weight from the ZnS film. The real-space density of states of the upper-surface TSS at the Dirac point (Fig. 2d) also confirms this identification. Therefore, the ZnS/Bi$_2$Se$_3$ system can be regarded as the same as a bare Bi$_2$Se$_3$ film placed in the vacuum, in the sense that the ZnS film remains a topologically trivial CI just like the vacuum, and hence the TSS in essence stays put at the surface of the Bi$_2$Se$_3$.

The same analysis yields very different behaviors for the system of ZnSe/Bi$_2$Se$_3$, shown in Figs. 2e-h. Compared to the former case of ZnS/Bi$_2$Se$_3$, the upper-surface TSS now shifts downward in energy due to enhanced charge transfer from ZnSe to Bi$_2$Se$_3$, which can be qualitatively understood from the smaller work function of the ZnSe film than that of ZnS (see Table I and also Supplementary Section S2). However, the most interesting difference between the present system and the former one is that, from Fig. 2e, a considerable weight of the TSS is carried by the ZnSe film, which is also unambiguously confirmed by the real space density of states of the TSS in Fig. 2h. Because a TSS marks the boundary between topologically inequivalent phases of matter, the upward shift of the TSS to the surface of the ZnSe overlayer indicates that the (ZnSe/TI) heterostructure as an entirety becomes an expanded TI via the topological proximity effect. More specifically, the topologization of the CI overlayer is signified by the whole (CI/TI) system possessing the same and only TSS located on top of the CI surface, namely, at the global boundary between the topologically trivial vacuum and the topologically nontrivial and combined (CI/TI) system.

We now discuss the underlying physical origin of the contrasting behaviors of the TSS in ZnS/Bi$_2$Se$_3$ and ZnSe/Bi$_2$Se$_3$. Since the nontrivial topology of the electronic structure of $Bi_2Se_3$ is ascribed to its strong SOC and small bulk band gap, it is intuitive that these factors associated with the CI materials are also essential for the different topological proximity effects we just observed. Indeed, the bulk phase of ZnSe has a much larger SOC (0.12 eV compared to 0.02 eV) and smaller band gap (1.20 eV compared to 2.09 eV) than ZnS, and the trend is preserved for the single-layer films as well (see Table I and Supplementary Section S1), making ZnSe "more proximate" to a TI material property-wise. Thus the term "proximity effect" here carries a dual meaning of both spatial and property-wise closeness. To gain a deeper understanding of the roles played by these key parameters of the Zn$M$, we look into the band structures of the two heterostructures (Figs. 2a, e). In the ZnSe/$Bi_2Se_3$ case (Fig. 2e), one can see that the SOC-induced gap ($\Delta_{SO}$) within the valence band of the ZnSe is aligned with the energy range of the TSS, and the TSS is mainly derived from the lower branch of the SOC-split bands. Charge transfer or transfer doping[39] from the ZnSe to the $Bi_2Se_3$ down-shifts the TSS, thereby facilitating its hybridization with the lower branch of the SOC-split bands. In particular, when the energy range of the TSS of the TI substrate falls into the SOC-induced gap region of the ZnSe, the coupling between the initial TSS and the SOC-split bands of the CI will be strong, leading to the newly-derived TSS of the resultant (CI/TI) system that contains high electron weight from the ZnSe overlayer. Moreover, because the dispersion of the lower branch of the SOC-split band is similar to the linear dispersion of the TSS, these two bands are more likely to hybridize in a wider range of the $k$-space.

In contrast, for ZnS/$Bi_2Se_3$, the bulk gap (and correspondingly the TSS) of the TI falls completely within the gap of the CI film (see Figs. 2a-c); consequently, the bands from the ZnS and $Bi_2Se_3$ are only mutually weakly perturbed. This "bad" alignment in the sense of absent proximity effect is partly due to the much larger band gap of ZnS than that of ZnSe, and partly because of the smaller mismatch between the work functions of the ZnS film and $Bi_2Se_3$ substrate than that in the $ZnSe/Bi_2Se_3$ case (Table I). As an alternative view, the seemingly trivial case of $ZnS/Bi_2Se_3$ in essence represents a balanced outcome of more accurate tuning, where one or more of the key CI parameters are pushed out of the range for the topological phase transition to take place.

**Reverse topological proximity effect in $ZnTe/Bi_2Se_3$.** With these competing factors in mind, we next look into the system of $ZnTe/Bi_2Se_3$. Since ZnTe has an even larger SOC and a smaller bulk band gap than that of ZnSe (see Supplementary Fig. 2) [40], the first intuition is to see an even more pronounced relocation of the TSS to the top of the ZnTe film. However, as a counterintuitive surprise, the upper-surface TSS now has a reverse shift toward the inner of $Bi_2Se_3$, with its peaked density located on the top of the $2^{nd}$ QL (Figs. 3a-e).

To reveal the physical origin of this unexpected reverse-proximity effect, we first note that, from Figs. 3a-d, both the bulk bands of $Bi_2Se_3$ and the TSS have a more noticeable downward shift in energy compared to the case of $ZnSe/Bi_2Se_3$, while the ZnTe film becomes strongly *p*-doped, indicating the most pronounced charge transfer from the CI overlayer to the TI substrate among the three cases. This behavior can be understood from the much smaller work function of the ZnTe film in comparison to that of ZnSe. The concomitant misalignment between the TSS and valence band top of ZnTe prevents the ZnTe film from being topologized by the $Bi_2Se_3$ substrate, as we have discussed before. Moreover, because of the significant charge transfer, the binding energy between the ZnTe film and the $Bi_2Se_3$ substrate (Table I) is much larger (0.29 eV) than the other two systems (< 0.1 eV). On an atomic scale, for the heterostructures investigated in the present study, the cation (Zn) in a given Zn*M* overlayer is located directly above the anion (Se) of the Bi$_2$Se$_3$ substrate; therefore, the competition between the in-plane Zn-*M* bonding and the out-of-plane Zn-Se bonding will significantly influence the interfacial coupling strength. Specifically, as all the out-of-plane bonds in the systems of Zn*M*/Bi$_2$Se$_3$ are given by the same type of Zn-Se bond, variations in the interfacial coupling strength for the three different systems lie in the different in-plane Zn-*M* bonds. Qualitatively, we further observe that the magnitudes of the band gaps for Zn*M* (*M* = S, Se, Te) satisfy $V_g$(ZnS)>$V_g$(ZnSe)>$V_g$(ZnTe), suggesting that the bond strengths should also satisfy $V_b$(Zn-S)>$V_b$(Zn-Se)>$V_b$(Zn-Te). Therefore, the in-plane bonding in ZnTe/Bi$_2$Se$_3$ is the weakest, making the interfacial bonding in ZnTe/Bi$_2$Se$_3$ the strongest. Since the QLs of Bi$_2$Se$_3$ are mutually coupled through weak van der Waals-like interactions, such a strong coupling between the ZnTe film and the 1$^{st}$ QL will compete with and weaken the interaction between the 1$^{st}$ and 2$^{nd}$ QL. As a consequence, the ZnTe film forces the 1$^{st}$ QL of the TI to be electronically partially decoupled from the remaining QLs. Furthermore, because one QL of Bi$_2$Se$_3$ does not have gapless TSS (see Supplementary Section S3)[41], the upper-surface TSS will naturally be relocated to the top of the 2$^{nd}$ QL. In other words, the CI has prevailed by topologically "trivializing" the 1$^{st}$ QL of the TI via the reverse-proximity effect.

To provide further support of the physical picture revealed above, we have performed two additional comparative studies. In the first one, we kept the relative positions of the ZnTe film and the 1$^{st}$ QL of Bi$_2$Se$_3$ fixed, but increased the distance between the 1$^{st}$ and 2$^{nd}$ QL from 2.5 Å to 3 Å, hence further decreasing their coupling. As shown in Figs. 3f-j, the TSS now becomes even more localized on top of the 2$^{nd}$ QL. In the second study, we fixed all the QLs of Bi$_2$Se$_3$ and slightly increased the distance between the ZnTe film and the 1$^{st}$ QL from 2.6 to 3 Å. As expected, the TSS now moves back to the top of the 1$^{st}$ QL, as shown in Figs. 3k-n.

As yet another strong and complementary support of the above picture, here we study the system of ZnTe/Bi$_2$Te$_3$, i.e., replacing Bi$_2$Se$_3$ with the structurally nearly identical Bi$_2$Te$_3$, which is also a TI. The charge transfer is less dramatic than that in ZnTe/Bi$_2$Se$_3$, as shown in Figs. 3p-s. The reason lies in the smaller work function mismatch between the tensilely strained ZnTe and Bi$_2$Te$_3$ than that between the compressively strained ZnTe and Bi$_2$Se$_3$ (Table I). The band gap of the ZnTe film is also decreased from 1.85 eV in the ZnTe/Bi$_2$Se$_3$ system to 1.39 eV in ZnTe/Bi$_2$Te$_3$, because of the different strains in the two systems (Table I). As a consequence, both the band misalignment (Fig. 3p) and binding energy (Table I) between the ZnTe film and the Bi$_2$Te$_3$ substrate become smaller compared to that in the former case. Naturally, the TSS now floats to the top of the ZnTe film, indicating that the ZnTe is now topologized by the Bi$_2$Te$_3$ substrate.

For the specific system of ZnTe/Bi$_2$Se$_3$, the interfacial coupling strength is fixed under normal contact conditions. Nevertheless, when an electric bias potential is applied in the direction perpendicular to the interface, the resultant charge transfer across the interface can be altered, potentially weakening the overall interfacial coupling to be comparable to the case of ZnTe/Bi$_2$Te$_3$ or ZnSe/Bi$_2$Se$_3$. If so, the reverse topological proximity effect is likely to be tuned into a normal proximity effect. This intriguing feasibility will be explored systematically in a future study.

It is also instrumental to elaborate more on the delicate roles of the different physical parameters influencing the dual proximity effects and the corresponding topological phase transitions. First, the overall coupling strength at the CI/TI interface, measured by the interfacial binding energy ($E_b$), is naturally a crucial parameter. This strength is largely determined by the relative work functions of the CI and TI (which in turn depend on their relative band gaps), the resultant charge transfer across the interface, and the nature of the local interfacial bonding. In particular, for the systems considered here, we observe that, (1) the direction of charge transfer is always from the cation (Zn) of the CI to the anion (Se) of the TI and, (2) stronger interfacial coupling exists for heterostructures in which the CI has a smaller work function than the TI. To see how the magnitude of $E_b$ is closely correlated with the topological proximity effects, we further observe that, when the CI/TI interfacial coupling is extremely weak (i.e., $E_b = 0.06$ eV for the given supercell size adopted) the location of the TSS will stay put at the interface, indicating absence of the topological proximity effect (Figs. 2a-d and Fig. 4a). In contrast, for ZnSe/Bi$_2$Se$_3$, the work functions of the CI and TI are nearly equivalent, the interfacial coupling strength is moderate (i.e., $0.06$ eV $< E_b < 0.20$ eV, see Figs. 2e-h, 3p-t, and 4b,c), and the topological proximity effect is sufficiently effective in inducing the topological phase transition as observed. Furthermore, for ZnTe/Bi$_2$Se$_3$, the work function of the CI is smaller than that of the TI, and the strong interfacial coupling is jointly reflected by the large binding energy (i.e., $E_b = 0.29$ eV) and the contaminant alignment of the TSS of the TI and the valence band of the CI. As a consequence, the substantial electronic hybridization at the interface will pull the first QL of the TI substrate to be even more weakly coupled with the second QL and the rest of the TI substrate, thereby topologically trivializing the first QL, and shifting the TSS to be on top of the second QL (Figs. 3a-e and 4d).

Aside from the interfacial coupling strength $E_b$, another delicate physical parameter in influencing the topological proximity effect is the spin-orbit coupling strength within the CI, $\Delta_{SO}$. This aspect can be qualitatively understood by recalling that the global topological property of the TI substrate is intimately tied to the spin-orbit coupling within the substrate[1]. Therefore, proper spin texture[30] of the CI overlayer due to its own spin-orbit coupling is also required in order for the topological proximity effect to be operative in the form of advancing the TSS from initially on the top of the TI to the top of the CI. This is qualitatively illustrated in Figs. 4b and 4c, showing that, when $\Delta_{SO}$ is weaker than some critical value $\Delta_C$, the TSS is still located at the interface (Figs. 2a-d, Fig. 4b); but when $\Delta_{SO} > \Delta_C$, the TSS will move to a lower energy due to stronger interfacial coupling (Fig. 4c), and spatially shift to the top of the CI (Figs. 2e-h, 3p-t).

Finally, we note that the dual-proximity effects revealed in TI-based heterostructures imply an inherent property of topological quantum phase transitions that do not invoke symmetry breaking measured by a local order parameter. This new type of quantum phase transition induced by topological proximity effects may find broad applicability. Motivated by these findings from density functional theory (DFT) calculations, an ongoing effort is to develop a more generic theory to describe the rich topological phase transitions in various CI/TI heterostructures based on consideration of the dominant physical factors controlling the transitions. Furthermore, experimental efforts demonstrating such tunability in the vertical location of the TSS, and other related manifestations and opportunities enabled by this tunability, are expected to materialize. Appealing examples include exploiting the robust nature of the TSS on top of a conventional catalyst for enhanced surface catalysis, spatially separating the TSS from the bulk state of a TI for potential definitive conductivity measurement of the TSS, protecting the TSS from gas contamination in air[33], etc. Such advances will also likely enrich the functionality and potential technological applications of the TI-based heterostructures.

**Methods**

All the density functional calculations were carried out using the Vienna *ab* initio simulation package (VASP) [42] with projector-augmented-wave potentials [43] and the Perdew-Burke-Ernzerhof generalized gradient approximation [44] for exchange-correlation functional. The lattice constants of $Bi_2Se_3$ and $Bi_2Te_3$ were adopted from experiments. The generic $Bi_2Se_3$ and $Bi_2Te_3$ substrates were modeled by a slab of 32 atomic layers or 6 QLs. The semiconducting overlayers Zn*M* (111) (*M* = S, Se, Te) were epitaxially placed on the (0001) surface of the TIs, and the energetically most stable configuration is obtained by putting Zn directly above Se, and *M* in the *hcp* sites, with the Zn face adjacent to the $Bi_2Se_3$ surface (Fig. 1d). The vacuum layers used are over 20 Å thick to ensure decoupling between neighboring slabs. During structural relaxation, atoms in the TI substrate were fixed in their respective bulk positions, and all the other atoms in the semiconducting overlayer were allowed to relax until the forces on them are smaller than 0.01 eV/Å. Test calculations with the first QL of the TI substrates also fully relaxed show that the relaxation effects are minimal and do not change the main physical pictures emphasized in the present study. A 7×7×1 k-point mesh was used for the 1×1 surface unit cell[45]. The SOC splitting of a semiconductor is defined as $\Delta_{SO} = \varepsilon\left(\Gamma_{8v}\right) - \varepsilon\left(\Gamma_{7v}\right)$ at the top of the valence band [40]. For the single-layered Zn*M* films, we use the magnitude of the SOC-split gaps in the valence band as a measure of the SOC.

**ACKNOWLEDGMENTS.** The authors thank Dr. Haiping Lan and Dr. Wenguang Zhu for helpful discussions. This work was supported by the National Natural Science Foundation of China (Grant Nos. 11034006, 51074151, and 11204286), The MOST (Grant Nos. 2010CB923401 and 2011CB921801), US National Science Foundation (Grant No. 0906025), and the BES program of US Department of Energy (Grant No. ER45958). The calculations were performed at the National Energy Research Scientific Computing Center (NERSC) of the US Department of Energy.

**Author contributions**

Z.Z. conceived the idea and directed the project. G.W. performed the DFT calculations. G.W., X.C., and H.C. wrote the initial draft of the paper, which was further polished by X.L. and Z.Z.. All authors contributed to the interpretation of the results and conceptual development of the work.

**Additional information**

The authors declare no competing financial interests. Supplementary Information accompanies this paper at http://www.nature.com/nmat. Reprints and permission information is available online at http://www.nature.com/reprints.

**Figure legends**

**Figure 1 | Schematic illustration on tuning the vertical location of the topological surface states (TSS) as a topological insulator (TI) is covered with a layer of conventional insulator (CI). a,** TSS floating to the top of the CI. **b**, Staying put at the CI/TI interface. **c**, Diving into the TI. **d**, The atomic structure of Zn$M$/Bi$_2$Se$_3$ ($M$ = S, Se, Te). The red lines denote the TSS; the arrows indicate the resulting directions of the topological phase transition.

**Figure 2 | Band structures of ZnS/Bi$_2$Se$_3$ (upper row) and ZnSe/Bi$_2$Se$_3$ (lower row) along the K−Γ−M direction.** The dots indicate the electronic bands contributed by the CI (**a** and **e**), the 1$^{st}$ QL of the TI (**b** and **f**), and the 6$^{th}$ QL (**c** and **g**), respectively; the sizes and colors of the dots also indicate different spectral weights and contributions from different atoms, respectively. **d** and **h** show the charge density distribution of the upper-surface TSS at the Γ point marked by the circle and indicated by DP$_U$. The DP$_{U/L}$ stands for the Dirac point at the upper/lower surface. The grey and cyan bars denote the locations of the different QLs and the CI, respectively.

**Figure 3 | Band structures of ZnTe/Bi$_2$Se$_3$ and ZnTe/Bi$_2$Te$_3$ along the K−Γ−M direction.** Band structures of the relaxed ZnTe/Bi$_2$Se$_3$ system (top row), the ZnTe/Bi$_2$Se$_3$ system with an increased separation of 3 Å from 2.5 Å between the 1$^{st}$ QL and 2$^{nd}$ QL of the TI substrate (second row), the ZnTe/Bi$_2$Se$_3$ system with an increased separation of 3 Å from 2.6 Å between the CI and 1$^{st}$ QL (third row), and the relaxed ZnTe/Bi$_2$Te$_3$ system

(bottom row). All other symbols are the same as in Fig. 2.

**Figure 4 | Illustrations on the effects of the key physical parameters on topological phase transitions. a**, Weak interfacial coupling. b & **c**, Moderate interfacial coupling, but with different spin-orbit coupling strengths. **d**, Strong interfacial coupling. The red triangles denote the TSS; the solid and dotted black curves correspond to the valence bands of the CI and the Fermi level of the heterostructures, respectively.

**TABLE I | Characteristic properties and vertical locations of the TSS in different CI/TI heterostructures.** The values for the band gap ($V_g$), spin-orbit splitting ($\Delta_{SO}$), and work function ($\Phi$) are obtained for free-standing single-layered CIs; the bulk lattice constants ($a$) of the CI are from experiments. The last three columns indicate the lattice mismatch ($\Delta a$), binding energy ($E_b$) between the CI and TI, and the location of the TSS ($Z$). The values in the square brackets are the corresponding values for the TI substrates.

| System (Substrate) | $V_g$ (eV) | $\Delta_{SO}$ (eV) | $\Phi$ (eV) | $a$ (Å) | $\Delta a$ | $E_b$ (eV) | $Z$ |
|---|---|---|---|---|---|---|---|
| ZnS (Bi$_2$Se$_3$) | 1.82 (0.40) | 0.03 | 6.10 (5.55) | 3.82 (4.14) | +8.4% | 0.06 | Interface |
| ZnSe (Bi$_2$Se$_3$) | 1.58 (0.40) | 0.20 | 5.77 (5.55) | 4.00 (4.14) | +3.5% | 0.09 | Top |
| ZnTe (Bi$_2$Se$_3$) | 1.85 (0.40) | 0.55 | 3.98 (5.55) | 4.31 (4.14) | -3.9% | 0.29 | Inside |
| ZnTe (Bi$_2$Te$_3$) | 1.39 (0.39) | 0.48 | 4.93 (5.00) | 4.31 (4.38) | +1.6% | 0.19 | Top |

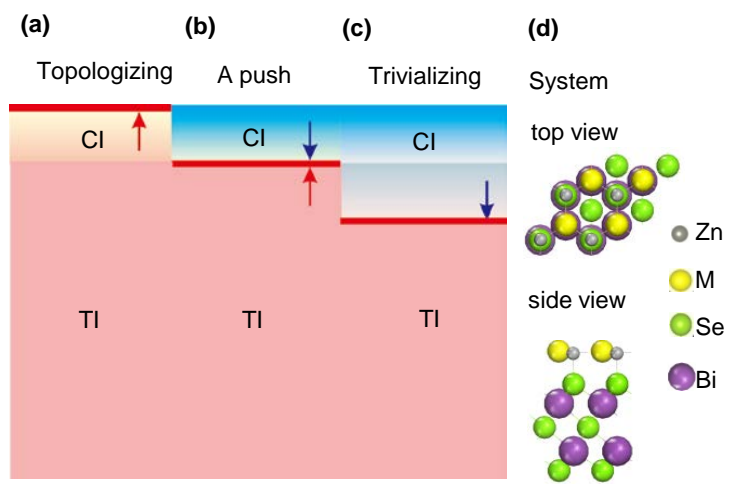

Fig. 1

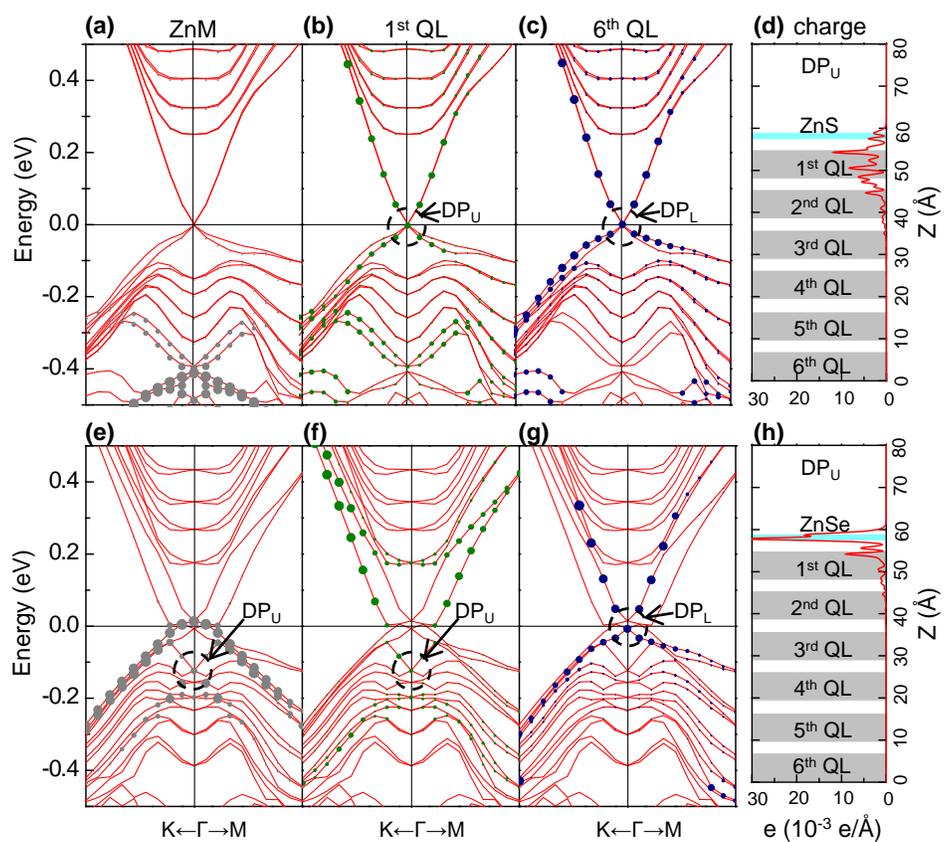

Fig. 2

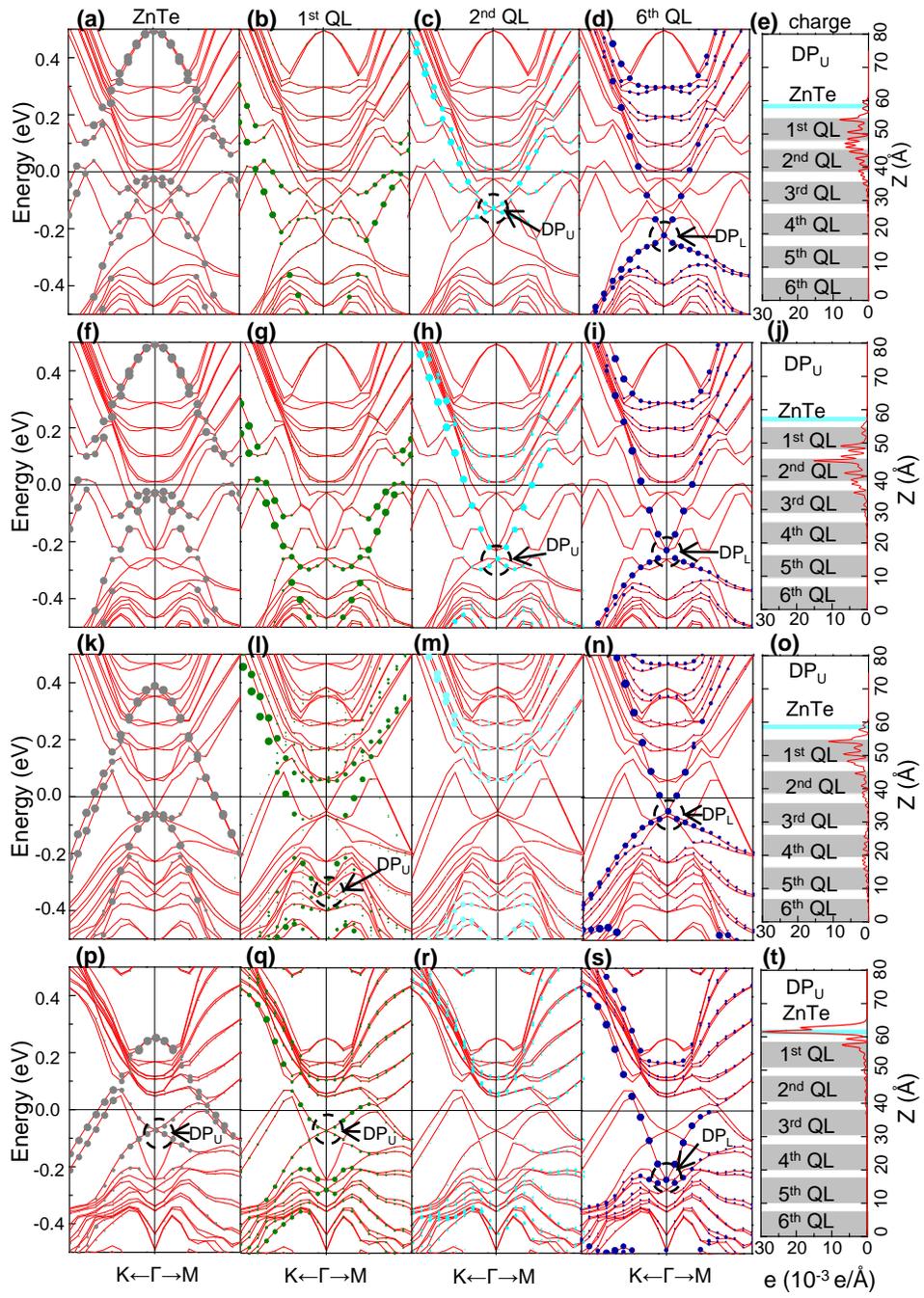

Fig. 3

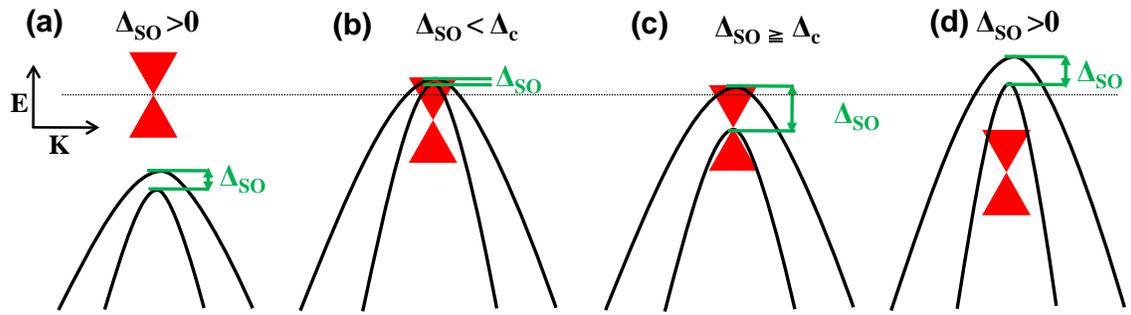

Fig. 4